\documentclass[preprint,showpacs,showkeys,preprintnumbers,nofootinbib,groupedaddress,superscriptaddress]{revtex4}

\usepackage[dvips]{graphicx}
\usepackage{color} 
\usepackage{epsfig}
\usepackage{fancybox}
\usepackage{amsmath}
\usepackage{amssymb}

\begin{document}

\title{Enhancement of lepton flavor violation in a model with bi-maximal
mixing at the grand unification scale} 


\author{Shinya Kanemura}
\email{kanemu@het.phys.sci.osaka-u.ac.jp}
\affiliation{Department of Physics, Osaka University, Toyonaka, Osaka
560-0043, Japan}

\author{Koichi Matsuda}
\email{matsuda@het.phys.sci.osaka-u.ac.jp}
\affiliation{Department of Physics, Osaka University, Toyonaka, Osaka
560-0043, Japan}

\author{Toshihiko Ota}
\email{toshi@het.phys.sci.osaka-u.ac.jp}
\affiliation{Department of Physics, Osaka University, Toyonaka, Osaka 560-0043, Japan}

\author{Tetsuo Shindou}
\email{shindou@sissa.it}
\affiliation{Scuola Internazionale
Superiore di Studi Avanzati, I-34014, Trieste, Italy}
\affiliation{Theory group, KEK, Tsukuba 305-0801, Japan \vspace{1cm}}

\author{Eiichi Takasugi}
\email{takasugi@het.phys.sci.osaka-u.ac.jp}
\affiliation{Department of Physics, Osaka University, Toyonaka, Osaka
560-0043, Japan}

\author{Koji Tsumura}
\email{ko2@het.phys.sci.osaka-u.ac.jp}
\affiliation{Department of Physics, Osaka University, Toyonaka, Osaka
560-0043, Japan}

\begin{abstract}
We study phenomenological predictions in the scenario with
the quasi-degenerate relation among neutrino Dirac masses, 
$m_{D1}^{} \simeq m_{D2}^{} < m_{D3}^{}$,
assuming the bi-maximal mixing at the grand unification scale
in supersymmetric standard models
with right-handed neutrinos.
A sufficient lepton number asymmetry can be produced 
for successful leptogenesis.
The lepton flavor violating process $\mu \to e \gamma$ can be
enhanced due to the Majorana phase, so that it can be  
detectable at forthcoming experiments.
The processes $\tau \to e \gamma$ and $\tau \to \mu \gamma$ 
are suppressed because of the structure of neutrino Dirac masses, 
and their branching ratios are smaller than that of $\mu \to e \gamma$.

%
\end{abstract}



\preprint{OU-HET 507}
\preprint{KEK-TH-993}
\preprint{SISSA  57/2005/EP}

\maketitle

\section{Introduction}
Neutrinos can be useful as a probe of physics at high energies 
such as grand unified theories (GUT).
Supersymmetry (SUSY) may be introduced to avoid problems
due to large hierarchy between the weak scale and the GUT scale.
Tiny neutrino masses and observed mixing angles may be explained 
by assuming the existence of right-handed neutrinos 
with large Majorana masses\cite{seesaw}.
They are determined from the high energy structure
of the model by using renormalization group equations (RGEs).
The resulting mass spectrum and mixing angles depend on
the Majorana mass matrix of right-handed neutrinos
and the neutrino Yukawa interaction. 
%
%
It would be possible to consider 
phenomenology of the model by putting additional 
assumptions in the high-energy structure 
of neutrino sector. 

In this paper, we consider the 
minimal supersymmetric standard model with 
right-handed neutrinos (MSSMRN),
in which the bi-maximal solution for 
the Pontecorvo-Maki-Nakagawa-Sakata (PMNS) matrix\cite{mns} 
is assumed to be realized at the GUT scale.
This solution is predicted in several GUT models\cite{model}.
In the model with the bi-maximal mixing solution,
the 1-3 element of the PMNS matrix is zero at the GUT scale,
while there are two Majorana CP phases\cite{majorana-cp}.
A non-zero value of the 1-3 element with the CP violating Dirac phase
can be induced at the low energy through RGEs\cite{cein,mst2}.
The observed value $\theta_{\odot}$ 
($\tan^2\theta_{\odot}\simeq 0.4$\cite{exp-sk,exp-sno,exp-kam}) 
for the solar neutrino angle at low energy is clearly
different from the maximal mixing $\pi/4$ at the GUT scale.
This difference can be explained by taking into account
the running effect due to the neutrino Yukawa couplings 
between the two scales\cite{cein,mst1,mst2,aklr}.
When masses of neutrinos corresponding 
to the solar neutrino data are relatively larger such as 0.05eV, 
the running effect becomes significant so that the value of 
$\theta_{\odot}$ can be reproduced at the low energy scale\cite{mst1,mst2}.
On the other hand, when the mass scale of neutrino is larger than
0.15eV, 
the atmospheric neutrino mixing angle is so instable that 
the bi-maximal mixing model cannot explain the experimental
result of atmospheric neutrino oscillation\cite{mst1,mst2}.
Therefore, we here consider the case in which masses of neutrinos 
are in the range between 0.05eV and 0.15eV.
Then the solar neutrino data prefer two 
cases for the pattern of the eigenvalues of 
the neutrino Dirac mass matrix;
(i) hierarchical case ($m_{D1}^{} < m_{D2}^{} < m_{D3}^{}$) and  
(ii) quasi-degenerate case ($m_{D1}^{} \simeq m_{D2}^{} < m_{D3}^{}$).
The case (i) has been studied in Ref.~\cite{st}, and it has been found that
lepton flavor violating processes are not significant.
In the present paper, we study the case (ii) and investigate its low energy
phenomenology. 

We shall show that our scenario is compatible with the low energy neutrino data, 
and that 
sufficient amount of lepton number asymmetry can be produced
for successful leptogenesis\cite{eps-leptogene,lepto-m,
resonant-lepto,resonant-lepto-1,leptogenesis-2}. 
Furthermore, we find that the lepton flavor violating process 
$\mu \to e \gamma$ can be enhanced by the Majorana phase effect
to be as large as the experimental reach at MEG\cite{MEG}.
We also find that the branching ratios of $\tau \rightarrow e \gamma$
and $\tau \rightarrow \mu \gamma$ are smaller than that of $\mu \to e \gamma$.
These are striking features of the quasi-degenerate scenario.

In Sec.2, the quasi-degenerate scenario with the bi-maximal mixing
solution is defined in the MSSMRN.
In Sec.3, we discuss phenomenological results of our scenario, especially 
on leptogenesis and lepton flavor violation.
Comments and conclusions are given in Sec.4.
Some derivations are given in Appendices.

\section{The quasi-degenerate scenario}
\label{sec_rg}

We consider the neutrino Yukawa couplings 
in the quasi-degenerate scenario
in the MSSMRN. 
The Lagrangian relevant to right-handed neutrinos is given by
\begin{align}
\mathcal{L}_{Y+M}=
\overline{N}_R \phi_u^{0} Y_{\nu} \nu_L^{}
-\frac{1}{2} \overline{N}_R^{c} M_{R} N_R 
+ \text{h.c.},
\end{align}
where $N_{R}$ is the right-handed neutrino 
with the $3\times 3$ Majorana mass matrix $M_{R}$, 
$\nu_{L}^{}$ is the left-handed neutrino, 
$\phi_{u}^{0}$ ($\phi_{d}^{0}$) is the neutral component of the Higgs doublet 
with the hypercharge $-1/2$ ($+1/2$), 
and 
$Y_{\nu}$ is the  $3\times 3$ Yukawa matrix for the neutrinos. 
The left-handed neutrino mass matrix is expressed 
at each scale as\cite{seesaw} 
\begin{align}
m_{\nu}
=
\frac{v^{2} \sin^{2} \beta}{2}
Y_{\nu}^T M_{R}^{-1} Y_{\nu},
\label{eq:seesaw}
\end{align}
where the vacuum expectation values $\langle \phi_{u}^{0} \rangle$ and
$\langle \phi_{d}^{0} \rangle$ satisfy
$v  = \sqrt{2} \sqrt{ \langle \phi_{u}^{0} \rangle^{2} + \langle
\phi_{d}^{0} \rangle^{2} } \simeq 246$ GeV 
and $\tan \beta = \langle \phi_{u}^{0} \rangle / \langle \phi_{d}^{0} \rangle$.

Let us consider the neutrino mass matrix at the GUT scale, $M_{X}$,
which is much higher than that of the Majorana masses of right-handed neutrinos.
We take the basis such that the mass
matrix of right-handed neutrinos is diagonal as
$ M_{R}=D_R^{}\equiv\mathrm{diag}(M_1,M_2,M_3)$,
where $M_i$ 
are real positive eigenvalues ($M_1 \leq M_2 \leq M_{3}$),
and that the mass matrix of the charged leptons is also diagonal.
The neutrino Dirac mass matrix $m_D^{}$ is diagonalized as
\begin{align}
m_D^{} 
\equiv Y_{\nu}\frac{v\sin\beta}{\sqrt{2}} 
= V_R^{\dagger} D_D^{} V_L\;,
\label{vr-dd-vl}
\end{align}
where $D_D^{}$ is a
diagonal matrix $D_D^{}\equiv\mathrm{diag}(m_{D1}^{},m_{D2}^{},m_{D3}^{})$ 
with real positive eigenvalues $m_{Di}^{}$ $(m_{D1}^{}\leq m_{D2}^{}\leq m_{D3}^{})$,
and $V_R$ and $V_L$ are unitary matrices.
As an important assumption of our model, 
we suppose that the neutrino mass matrix satisfies 
the bi-maximal mixing solution at $M_X$; i.e., 
\begin{align}
m_{\nu}(M_X)=O_B D_{\nu}O_B^T\;,
\label{eq:mNuatGUT}
\end{align}
where $O_B$ is given by
\begin{align}
O_B\equiv
\begin{pmatrix}
\frac{1}{\sqrt{2}}&-\frac{1}{\sqrt{2}}&0\\
\frac{1}{2}&\frac{1}{2}&-\frac{1}{\sqrt{2}}\\
\frac{1}{2}&\frac{1}{2}&\frac{1}{\sqrt{2}}
\end{pmatrix}\;,
\end{align}
and $D_{\nu}$ is a diagonal matrix,
\begin{align}
D_{\nu} \equiv \mathrm{diag}(m_1, m_2 e^{i\alpha_0}, m_3 e^{i\beta_0}),
\end{align}
with $\alpha_0$ and $\beta_0$ being the Majorana phases 
and $m_{i}$ being real positive\cite{majorana-cp}.

It is known that when the scale of neutrino masses are so large as
$0.05\text{eV}<m_1\sim m_2\equiv m$,
the running effect on the neutrino mass matrix 
between the weak scale $m_{Z}^{}$ and $M_{X}$
becomes large 
due to the neutrino Yukawa interaction\cite{rge-sol, mst1, mst2}.
The 1-3 element $V_{13}$ of the PMNS matrix is also induced 
at the low energy scale,  
which is found to be
proportional to $m_{1} m_{3}$\cite{mst2}.
The element $|V_{13}|$ can be sizable when both $m_{1}$ and
$m_{3}$ are sufficiently large\footnote{%
There is also a chance to appear large running effect in the case where
the neutrino mass spectrum is inverse hierarchical, i.e., $m_{3}\ll
m_{1} < m_{2}$.
In this case, different prediction for the 1-3 element of the PMNS
matrix is obtained at the low energy scale.
}.
In order to reproduce the solar neutrino data 
from the bi-maximal solution at $M_{X}$
with $Y_{\nu}^{\dagger} Y_{\nu}$ to be diagonal,
there are two possibilities for the pattern of 
the neutrino Dirac masses,
i.e., hierarchical case $m_{D1}< m_{D2} < m_{D3}$ and quasi-degenerate
case $m_{D1} \simeq m_{D2} < m_{D3}$ with 
\begin{align}
V_L=P_{ex}\equiv
\begin{pmatrix}
0&0&1\\
0&1&0\\
-1&0&0
\end{pmatrix}\;.
\label{pex}
\end{align}
The detailed discussion appears in Appendix A.
In this paper, we concentrate on the latter case, namely the quasi-degenerate
case.
The hierarchical case has been studied in Ref.~\cite{st}.

From Eqs.~(\ref{eq:seesaw}), (\ref{vr-dd-vl}), and (\ref{eq:mNuatGUT}), 
we obtain
\begin{align}
\widetilde{M}_R^{-1}\equiv (V_R^*D_R^{-1}V_R^{\dagger})
=D_D^{-1}(P_{ex}O_B)D_{\nu}(P_{ex}O_B)^TD_D^{-1}\;.
\label{mrinv}
\end{align}
The unitary matrix $V_{R}$ as well as the eigenvalues $M_{i}$ are obtained by diagonalizing 
$\widetilde{M}_R^{-1}$.
Consequently, we find
\begin{align}
V_R=
\begin{pmatrix}
\frac{1}{\sqrt{2}}& \frac{1}{\sqrt{2}}\cos\frac{\zeta}{2}
& - \frac{i}{\sqrt{2}}\sin\frac{\zeta}{2}\\
 -\frac{1}{\sqrt{2}}& \frac{1}{\sqrt{2}}\cos\frac{\zeta}{2}
& -\frac{i}{\sqrt{2}}\sin\frac{\zeta}{2}\\
0& -i\sin\frac{\zeta}{2}&\cos\frac{\zeta}{2}
\end{pmatrix}
\begin{pmatrix}
e^{-\frac{i}{2}\beta_{0}}&&\\
& e^{-\frac{i}{4}\alpha_{0}}&\\
&& e^{-\frac{i}{4}\alpha_{0}}
\end{pmatrix}\;,
\label{eq_vr}
\end{align}
where  
\begin{align}
\tan \zeta \equiv \frac{2r}{1+r^2}\tan\frac{\alpha_0}{2}, 
\quad \text{with} \quad
r\equiv m_{D1}^{}/m_{D3}^{}<1\;.
\label{eq:defzeta}
\end{align}
As seen in Eq. \eqref{eq_vr}, 
some off-diagonal elements of $V_{R}$ are of order one. 
This is the striking feature of the quasi-degenerate case 
in contrast with the hierarchical case
in which $V_{R}$ is approximately the unit matrix\cite{st}. 
For the masses of the right-handed neutrinos, 
we obtain 
\begin{align}
M_1=&\frac{m_{D1}^2}{m}\;,\nonumber\\
M_{2}=&\frac{m_{D1}^2}{m}\frac{2}
{\sqrt{[1-r^2]^2\cos^2\frac{\alpha_0}{2}+4r^2}
+[1-r^2]\cos\frac{\alpha_0}{2}}\;, \label{Mi-expression}\\
M_{3}=&\frac{m_{D1}^2}{m}\frac{2}
{\sqrt{[1-r^2]^2\cos^2\frac{\alpha_0}{2}+4r^2}
-[1-r^2]\cos\frac{\alpha_0}{2}}\;. \nonumber
\end{align}
The derivation of Eqs. \eqref{eq_vr} and \eqref{Mi-expression} 
is shown in Appendix B.

\section{The phenomenology}

In this section, phenomenological consequences of our scenario 
are studied.
The parameters of the neutrino sector at $M_{X}$ 
are related to the low energy observables 
by the RGEs (see Appendix A).
They are constrained from the data of solar and
atmospheric neutrino experiments.  
We here take the values
$\Delta m_{\odot}^2 \equiv 8.3\times 10^{-5}\text{ eV}^2$ and 
$\tan^{2} \theta_{\odot}  \equiv 0.4$
as the solar neutrino results\cite{exp-kam}, 
and 
$\Delta m_{\text{atm}}^2 \equiv 2.5\times 10^{-3}\text{ eV}^2$ and 
$\sin^{2} 2 \theta_{\text{atm}} \equiv 1.0$
as the atmospheric neutrino results\cite{exp-atm}. 
Throughout this paper, $M_{X}$ is assumed to be $2 \times 10^{16}$ GeV.
In the following, 
after the discussion on basic properties,
we examine the consistency with leptogenesis, and predict lepton flavor
violation (LFV).

\begin{figure}
\begin{tabular}{cc}
\includegraphics[scale=0.6]{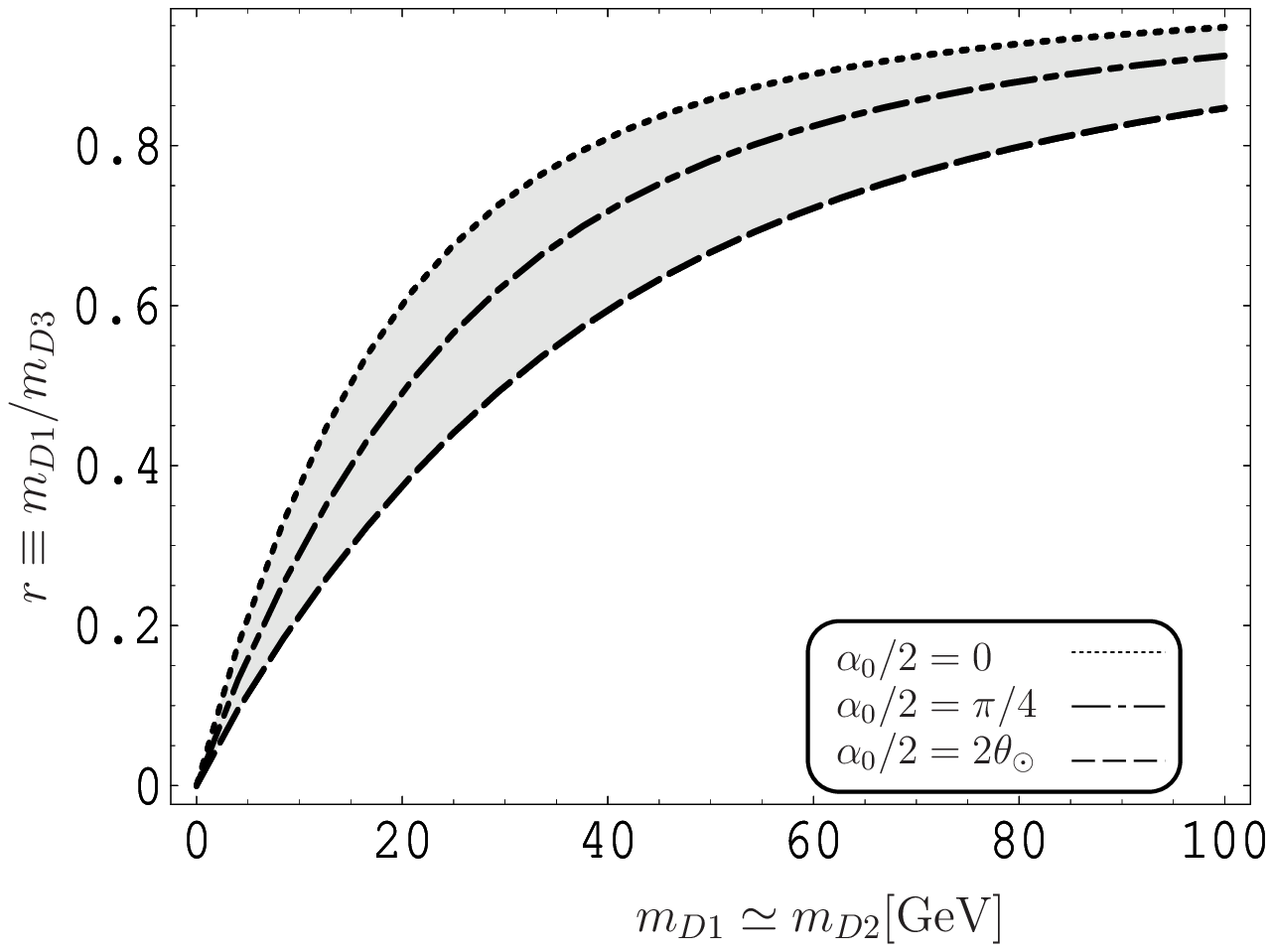}&
\includegraphics[scale=0.6]{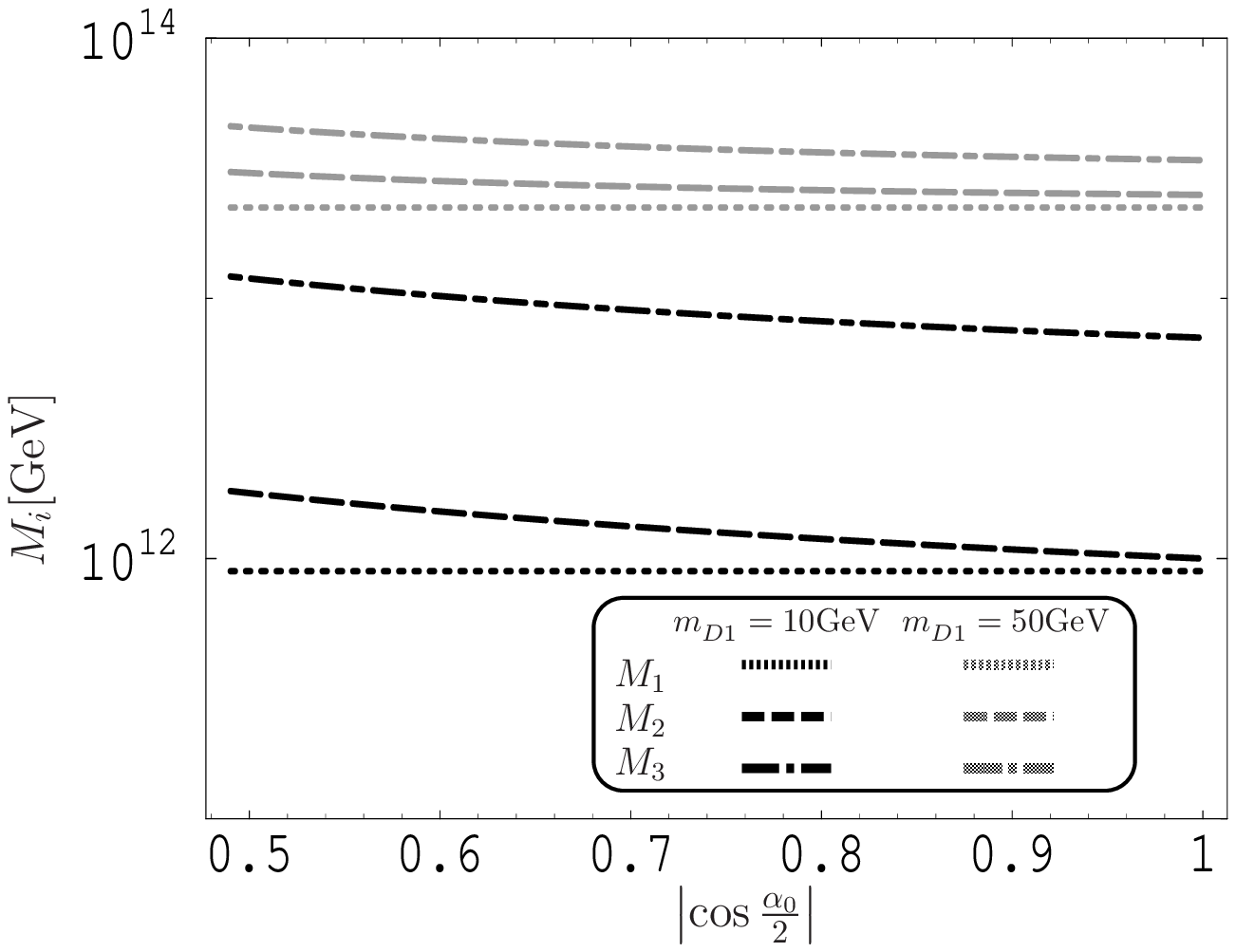}\\
(a)&(b)
\end{tabular}
\caption{(a)
The ratio $r$ as a function of $\alpha_0$ and $m_{D1}^{}$ for $\tan\beta=5$ and $m=0.1$ eV.
The shaded region corresponds to 
$\cos 2\theta_{\odot} \leq |\cos(\alpha_0/2)| \leq 1$.
(b) The eigenvalues $M_i$ as a function of $|\cos (\alpha_0/2)|$
for $m_{D1}^{}=10$ GeV (black curves) 
and $m_{D1}^{}=50$ GeV (gray curves)
for $\tan\beta=5$ and $m=0.1$ eV.
}
\label{Fig:mDvsr-and-MR}
\end{figure}

The experimental value $\theta_{\odot}$ can be reproduced
in our scenario with the bi-maximal solution at $M_{X}$. 
As we take the degenerate mass of neutrinos, the running effect
on the solar neutrino angle can be large between $M_{X}$ and $m_{Z}^{}$.
On the other hand, the angle 
$\theta_{\text{atm}}$ can be explained 
by assuming the masses of neutrinos are less than 0.15
eV so that the running effect is small\cite{mst1,mst2}.
In our scenario,
the running effects on the neutrino mass matrix,
which are parametrized by $\epsilon_e$ and $\epsilon_{\tau}$,
can be expressed as
\begin{align}
\epsilon_e =& -\frac{1}{8\pi^2}\frac{m_{D1}^2}{(v\sin\beta)^2}
              \left(\frac{1}{r^2}-1\right)\ln\frac{\overline{M}_R}{M_X}\;,\nonumber\\
\epsilon_{\tau} \simeq& 
 -\frac{1}{8 \pi^2} \frac{m_{\tau}^{2}}{(v \cos\beta)^{2}} \ln\frac{m_Z^{}}{M_X}\;.
\label{eq:epsilon-quasi-deg}
\end{align}
where $\overline{M}_{R}$ is the typical mass scale for right-handed
neutrinos and $m_{\tau}$ is the mass of the tau lepton: see Appendix A.
From  Eqs. \eqref{eq:epsilon-quasi-deg} and ~(\ref{eq:tanSqthetaSol}) with
the experimental data for angles and mass differences, 
we obtain the relation among $\alpha_0$, $m_{D1}^{}$ and $r$. 
In Fig.~\ref{Fig:mDvsr-and-MR}-(a), 
we show the ratio $r$ as a function of $m_{D1}^{}$ for each value of
$\alpha_{0}$ in the case of  $m=0.1$ eV and $\tan\beta=5$. 
We find that $r$ is insensitive to $\alpha_{0}$.
In Fig.~\ref{Fig:mDvsr-and-MR}-(b),
$M_i$ are shown
as a function of $|\cos (\alpha_0/2)|$
for $m_{D1}^{}=10$ GeV and $m_{D1}^{}=50$ GeV,
which are determined 
by $m$, $\alpha_0$, and $r$ through Eq.~(\ref{Mi-expression}).
Notice that $M_1$ and $M_2$ are coincident when $\alpha_0 \rightarrow0$
up to $\mathcal{O}(\Delta m_{\text{atm}}^{2}/m^{2})$.
The scale $\overline{M}_{R}$ takes the value between $10^{12}$ and $10^{14}$ GeV.
%
%

\subsection{Leptogenesis}

In models with the heavy Majorana neutrinos, 
it is possible to consider leptogenesis\cite{eps-leptogene} 
in order to explain a baryon asymmetry of the universe.
In leptogenesis, the out of equilibrium decays of heavy Majorana 
neutrinos produce a lepton number asymmetry which is converted
to the baryon number asymmetry through the sphaleron processes.
It is known that leptogenesis is successful to explain the 
baryon number of the universe when $m_3<0.15$eV and 
$M_1>2\times 10^7$GeV\cite{lepto-m}, and mass parameters
of our model can be in this allowed range.
On the other hand, a constraint from gravitino overproduction
can be a serious problem for our model. In order to allow 
the reheating temperature $T_R>10^{12}$GeV which is required for
case $M_1>10^{12}$GeV, the gravitino mass should be heavier than
10TeV\cite{gravitino}\footnote{In various SUSY models such as the minimal 
supergravity model, the gravitino mass are related to the soft SUSY 
parameters. In our paper, we don't touch origin of soft SUSY breaking
terms and we can take the gravitino mass as an 
independent parameter\cite{gravitino-1}, 
though assume the universal soft SUSY parameters.
}.

The lepton number asymmetry is produced in the decay 
of heavy Majorana neutrinos, which can be expressed as\cite{eps-leptogene}
\begin{align}
\epsilon_i
\simeq-\frac{1}{8\pi}
\sum_{k\neq i}f\left(\frac{M_k^2}{M_i^2}\right)
\frac{\mathrm{Im}[(Y_{\nu} Y_{\nu}^{\dagger})^2_{ik}]}
{(Y_{\nu} Y_{\nu}^{\dagger})_{ii}},
\label{eq_eps-lepto}
\end{align}
where
$f(x)$ is given in the MSSMRN as
\begin{align}
f(x)=\sqrt{x}\left[\frac{2}{1-x}+\ln\left(\frac{1+x}{x}\right)\right]\;.
\label{eq:fx}
\end{align}

In our scenario,
$(Y_{\nu} Y_{\nu}^{\dagger})_{ij}$ are calculated at the leading
order as 
\begin{align}
&(Y_{\nu} Y_{\nu}^{\dagger})_{ij} = \frac{2}{v^{2} \sin^{2} \beta} 
(V_R^{\dagger}D_D^2V_R)_{ij}
\nonumber\\
&\simeq 
\frac{2 m_{D1}^2}{v^{2} \sin^{2} \beta} \times \nonumber  \\
&
\begin{pmatrix}
1&-\frac{1}{2}\delta_1\cos\frac{\zeta}{2} e^{\frac{i}{4}(2\beta_0-\alpha_0)}&
\frac{i}{2}\delta_1\sin\frac{\zeta}{2}e^{\frac{i}{4}(2\beta_0-\alpha_0)}\\
-\frac{1}{2}\delta_1\cos\frac{\zeta}{2} e^{-\frac{i}{4}(2\beta_0-\alpha_0)}
&\frac{1}{2}(1\!+\!1/r^{2}\!+\!(1\!-\!1/r^{2})\cos\zeta)&
\frac{-i}{2}(1-1/r^{2})\sin\zeta \\
\frac{-i}{2}\delta_1\sin\frac{\zeta}{2}e^{-\frac{i}{4}(2\beta_0-\alpha_0)}&
\frac{i}{2}(1-1/r^{2})\sin\zeta&
\frac{1}{2}(1\!+\!1/r^{2}\!-\!(1\!-\!1/r^{2})\cos\zeta)
\end{pmatrix}\;,
\label{yydag}
\end{align}
where $\delta_1\equiv m_{D2}^{2} / m_{D1}^2 -1$ $(\ll 1)$.
The asymmetry $\epsilon_{3}$ is negligibly smaller than $\epsilon_{1,2}$
because of $M_{1} \sim M_{2} \ll M_{3}$.~\footnote{%
In the limit of $x \rightarrow 1$,
there is an enhancement effect 
in the lepton number
asymmetry\cite{resonant-lepto,resonant-lepto-1}. 
The enhancement is smeared by the following two reasons; i.e.,
(i) the effect of the decay width for $N_{i}$ 
and 
(ii) the effect of the small mass difference between $M_{1}$ and $M_{2}$
$\sim \mathcal{O}(\Delta m_{\text{atm}}^2/m^2)$.
} 
In the same reason, 
the elements including $(Y_{\nu} Y_{\nu}^{\dagger})_{12,21}$ 
are dominant in $\epsilon_{1,2}$.
Therefore $\epsilon_{1,2}$ are approximately proportional to
$\delta_1^2$, 
which means that finite $\epsilon_{1,2}$ appear only when
there is a deviation from the exact mass degeneracy $M_1=M_2$.

\begin{figure}
\begin{center}
\includegraphics{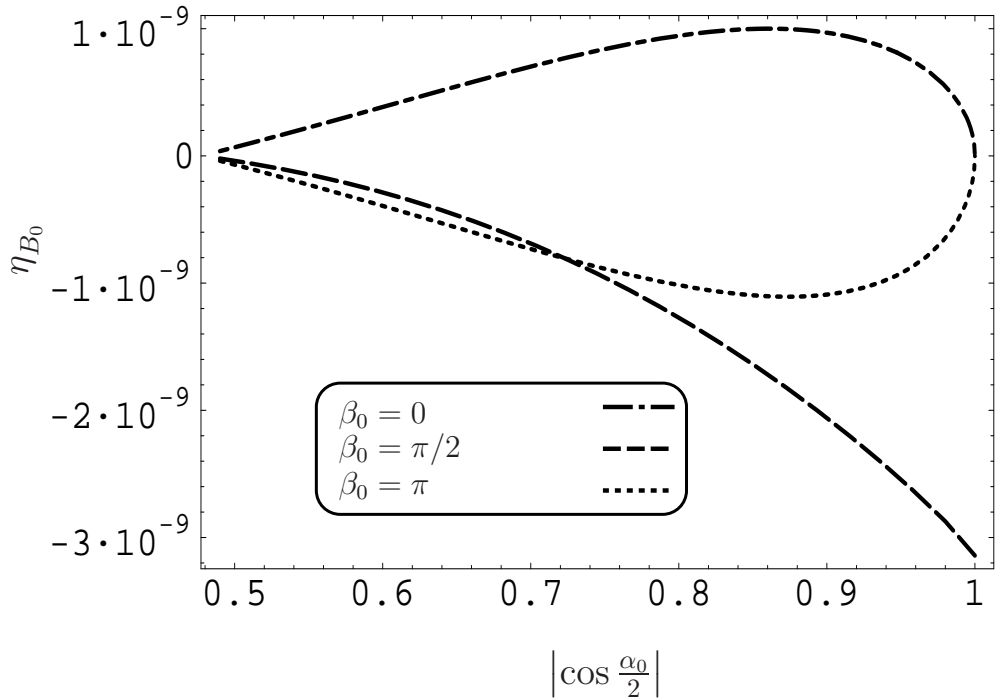}
\caption{The baryon number asymmetry $\eta_{B_{0}}^{}$ as a
function of $|\cos(\alpha_0/2)|$ for $\beta_0=0$, $\pi/2$ and $\pi$.
The other parameters are assumed to be
$m=0.1$ eV, $m_{D1}^{}=50$ GeV,
$\delta_1 =0.05$
and $\tan\beta=5$.}
\label{fig_bau}
\end{center}
\end{figure}

In leptogenesis, the baryon number asymmetry $\eta_{B_{0}}^{}$
of the Universe is
explained by using the lepton number asymmetry\cite{leptogenesis-2};
\begin{align}
\eta_{B_{0}}^{}\simeq -10^{-2}\kappa_0\sum_i \epsilon_i\;,
\end{align}
where $\kappa_0\simeq 0.3/\{ K(\ln K)^{3/5} \}$ with
$K\simeq 170 (m / \text{[eV]})$. 
The numerical result of $\eta_{B_{0}}^{}$ is given in Fig.~\ref{fig_bau}.
The experimental value $\eta_{B_{0}}^{}\sim 6.5\times 10^{-10}$
\cite{Spergel:2003cb} can be 
realized in our scenario.

\subsection{LFV processes}

In the MSSMRN, 
the slepton mixing can be a source of LFV\cite{lfv,real-y}.
Assuming the universal soft-breaking parameters at $M_{X}$, 
mixing among left-handed sleptons is induced 
by the renormalization group effects due to neutrino Yukawa couplings
between $M_X$ and $\overline{M}_R$,
even when there is no mixing at $M_{X}$.
The induced off-diagonal elements of the slepton mass matrix
are approximately expressed as\cite{lfv,real-y}
\begin{align}
(m_{\widetilde{L}}^2)_{ij}\simeq \frac{6m_0^2+2|A_0|^2}{16\pi^2}
(Y_{\nu}^{\dagger} \Omega Y_{\nu})_{ij}\quad (i\neq j)\;,
\label{eq:YnudagOmegaYnu}
\end{align}
where $m_{0}$ and $A_{0}$ are the universal SUSY breaking parameters, and 
\begin{align}
\Omega &\equiv
\mathrm{diag}\left(
\ln \frac{M_1}{M_X},\ln \frac{M_2}{M_X}, \ln
\frac{M_3}{M_X} \right).
\end{align}
These off-diagonal elements contribute to LFV processes
such as $\ell_i\to \ell_j\gamma\; (i\neq j)$. 
The decay widths are given by
\begin{align}
\Gamma(\ell_i\to \ell_j\gamma)\simeq& \frac{\alpha^3m_{\ell_i}^5}{192\pi^3}
\frac{|(m_{\widetilde{L}}^2)_{ij}|^2}{m_S^8}\tan^2\beta,
\label{eq:LFVWidth}
\end{align}
where $\alpha$ is the fine structure constant, 
and $m_S^{}$ denotes the typical mass scale of SUSY particles.
In the case of universal soft terms, $m_S$ is approximately
evaluated in Ref. \cite{ppty}.

Let us consider Eq.~\eqref{eq:YnudagOmegaYnu} in our scenario.
We can express $\Omega$ as 
\begin{align}
Y_{\nu}^{\dagger} \Omega Y_{\nu}
=
Y_{\nu}^{\dagger} 
\left\{
\ln \frac{M_{X}}{M_{3}} \ \pmb{1}
-
\mathrm{diag}
\left(\ln\frac{M_1}{M_3},\ln\frac{M_2}{M_3},0
\right) 
\right\} Y_{\nu}\;,
\label{separate_mdmd}
\end{align}
where
the second term of RHS in Eq.~(\ref{separate_mdmd})
corresponds to the threshold effect
of right-handed neutrinos. 
It has been often considered the case 
in which the first term  
gives dominant contributions to $\ell_i\to \ell_j\gamma$ processes.
However, $Y_{\nu}^{\dagger} Y_{\nu}$ is diagonal at $M_{X}$
so that the first term does not contribute to LFV. 
Therefore, remaining sources for LFV are 
in the second term of Eq.~(\ref{separate_mdmd}).
The off-diagonal elements of $(Y_{\nu}^{\dagger}\Omega Y_{\nu})_{ij}$
$(i\neq j)$
are found to be
\begin{align}
\left| (Y_{\nu}^{\dagger} \Omega Y_{\nu})_{12} \right|=&
\frac{m_{D2}m_{D3}}{\sqrt{2} v^{2} \sin^{2} \beta}\sin 2\zeta
 \ln\frac{M_2}{M_3}\;,
\label{eq:YnudagOmegaYnu12}\\
\left| (Y_{\nu}^{\dagger} \Omega Y_{\nu})_{13} \right|=&
\frac{m_{D1}m_{D3}}{\sqrt{2} v^{2} \sin^{2} \beta}\sin 2\zeta
 \ln\frac{M_2}{M_3}\;,
\label{eq:YnudagOmegaYnu13}\\
\left| (Y_{\nu}^{\dagger} \Omega Y_{\nu})_{23} \right|=&
\frac{m_{D1}m_{D2}}{v^{2} \sin^{2} \beta}
 \left(\ln\frac{M_1}{M_3}
 -\cos^2\frac{\zeta}{2}\ln\frac{M_2}{M_3}
 \right)\;.\label{eq:YnudagOmegaYnu23}
\end{align}

\begin{figure}
\includegraphics{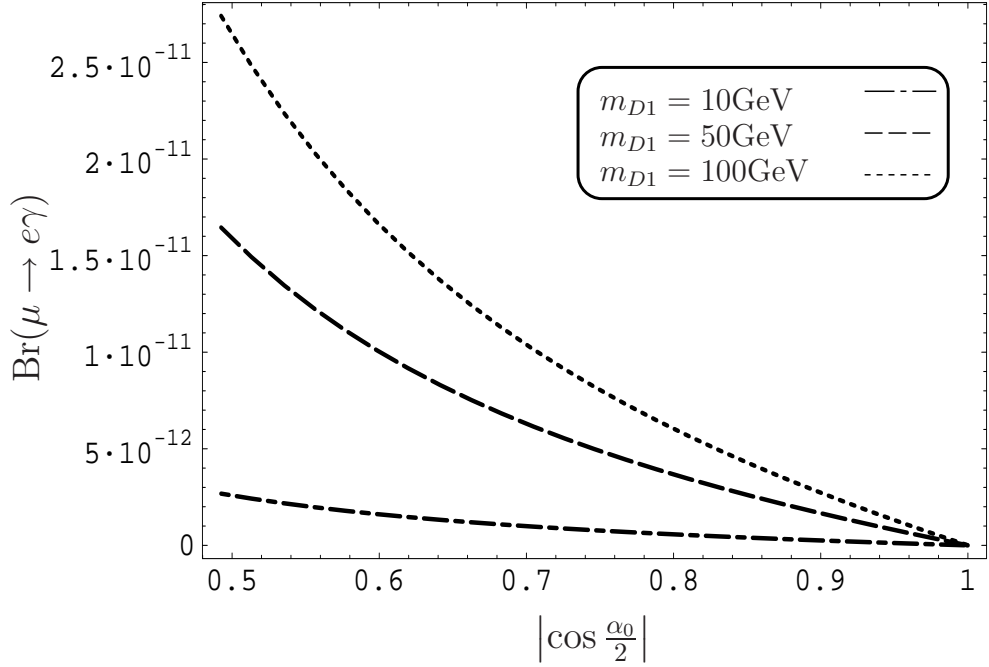}
\caption{The branching ratio of $\mu\to e\gamma$ as a function of
$|\cos (\alpha_0/2)|$ for $m_{D1}^{}=10$, $50$ and $100$ GeV.
The SUSY parameters are taken to be
$\tan\beta=5$, $m_0=200$ GeV, $A_0=100$ GeV and $m_S^{}=200$ GeV.}
\label{fig_muegamma}
\end{figure}

\begin{figure}
\includegraphics{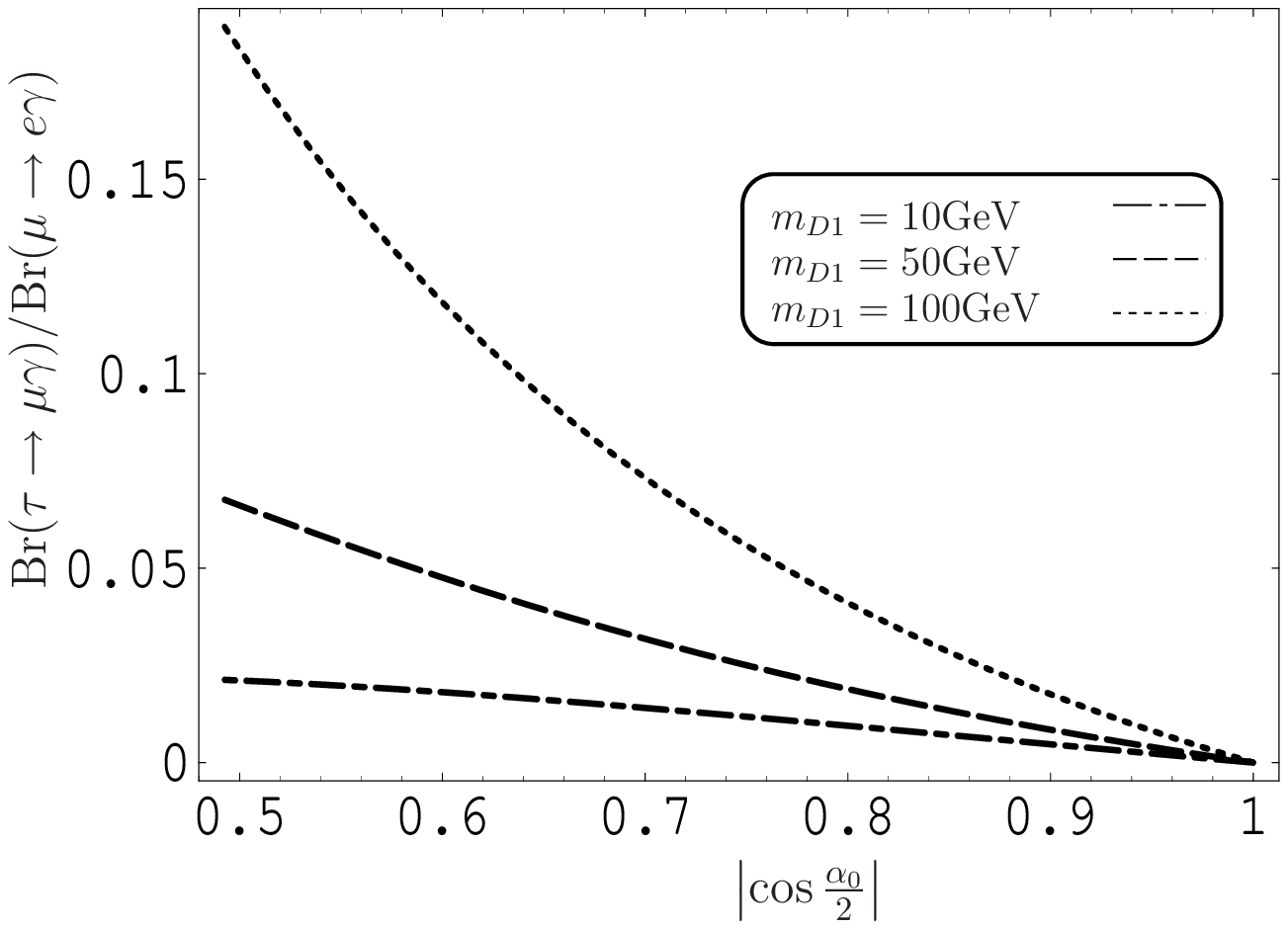}
\caption{The ratio of $\text{Br}(\tau\to \mu\gamma)$ to 
$\text{Br}(\mu\to e\gamma)$ as a function of $|\cos(\alpha_0/2)|$ for
$m_{D1}^{}=10$, $50$ and $100$ GeV. 
We take $\tan\beta=5$.}
\label{fig_rat23to12}
\end{figure}

The branching ratio of $\tau\to \mu \gamma$ 
is found to be smaller than those of $\mu \to e \gamma$ and $\tau \to e \gamma$ 
by a factor of $r$. 
All the processes $\ell_i\to \ell_j\gamma$ are maximally suppressed
for $|\cos (\alpha_0/2)| \rightarrow 1$ because of $M_1 \simeq M_2$.
From $m_{D1}^{} \simeq m_{D2}^{}$, 
we have the relation among the branching ratios of the LFV processes as
\begin{align}
\text{Br}(\tau\to e\gamma)\simeq 
\text{Br}(\tau\to\bar{\nu}_{e}\nu_{\tau} e)\text{Br}(\mu\to e\gamma)\;,
\end{align}
where Eqs.~\eqref{eq:LFVWidth}, \eqref{eq:YnudagOmegaYnu12}  
and \eqref{eq:YnudagOmegaYnu13} are used.
In Fig.~\ref{fig_muegamma},
we show $\text{Br}(\mu\to e\gamma)$ as a function of 
$|\cos (\alpha_0/2)|$.
We find that the value can reach $10^{-12}$ for the smallest value of
$|\cos (\alpha_{0}/2)|$ under $\cos 2 \theta_{\odot} \leq |\cos(\alpha_{0}/2)|
\leq1 $ 
and for $m_{D1}^{} \gtrsim 50$ GeV.
In the case of quasi-degenerate light neutrinos, it is known that
the lepton flavor violation processes are strongly suppressed 
with trivial right-handed mixings,{\em i.e.} the degenerate 
heavy neutrino mass spectrum and real mixings among right-handed 
neutrinos\cite{real-y}. However, Pascoli et al. showed the possibility
of enhancement of the lepton flavor violation processes due to the
existence of new CP phases in right-handed mixing even for
the case where both light and heavy neutrinos are 
quasi-degenerate\cite{complex-y}.
In our model, we don't introduce any CP phases other than $\alpha$ and $\beta$,
but the lepton flavor violation processes are 
nevertheless enhanced because of the threshold effects of heavy neutrinos.
Therefore we expect that current\cite{MEG} and future experiments can 
test our scenario 
through the LFV measurement for $\mu\to e \gamma$.
As compared to the hierarchical case, 
larger branching ratios for the LFV processes can be obtained 
in the quasi-degenerate case 
because of the mixing among right-handed neutrinos. 
In Fig.~\ref{fig_rat23to12}, we show the ratio of 
$\text{Br}(\tau\to \mu \gamma)$ to $\text{Br}(\mu\to e\gamma)$ as
a function of $|\cos(\alpha_0/2)|$.
In a wide range of the parameter space, $\text{Br}(\tau\to \mu\gamma)$ is smaller 
than $\text{Br}(\mu\to e\gamma)$. 
This is a striking feature of our scenario.

\section{Conclusion}

We have studied the quasi-degenerate scenario in the model in which  
the bi-maximal mixing solution is realized at $M_{X}$ in the MSSMRN.
By using the low energy neutrino data,
we have shown that our result is consistent with the WMAP data 
assuming leptogenesis.
Furthermore, it has been found that the LFV process $\mu \to e \gamma$ can 
be large enough to be detected at forthcoming experiments such as MEG.
We also have found that the branching ratios of $\tau \rightarrow e \gamma$
and $\tau \rightarrow \mu \gamma$ are smaller than that of $\mu \to e \gamma$.
The prediction of sizable LFV processes is a discriminative feature 
of the quasi-degenerate scenario as compared to the hierarchical one.   

It should be noted that our results strongly depend on the Majorana phases $\alpha_{0}$
and $\beta_{0}$.
In particular, the processes $\ell_{i} \to \ell_{j} \gamma$ 
become more enhanced for smaller values of
$|\cos (\alpha_{0}/2)|$.
We also note that 
the assumption of diagonal $Y_{\nu}^{\dagger} Y_{\nu}$ is crucial for
our results in the quasi-degenerate scenario.
The tau associated LFV processes might become significant 
when a non-zero 2-3 element of $Y_{\nu}^{\dagger}Y_{\nu}$ 
would be taken into account, by relaxing the assumption of diagonal
$Y_{\nu}^{\dagger} Y_{\nu}$.\\

\noindent
{\large \it Acknowledgments}\\
This work is supported in part by Japan Society for Promotion
of Science (Nos. 15-03693, 15-03700 and 15-03927), and also by
the Japanese Grant-in-Aid for Scientific Research of Ministry of
Education, Science, Sports and Culture, No. 12047218.\\

\appendix

\section{RGE analysis}
\numberwithin{equation}{section}

The RGE for the neutrino mass matrix is given by
\begin{align}
\frac{{\rm d} m_{\nu}}{{\rm d}\ln \mu}=
\frac{1}{16\pi^2}\left\{
[(Y_{\nu}^{\dagger}Y_{\nu})^T+(Y_e^{\dagger}Y_e)^T]m_{\nu}
+m_{\nu}[(Y_{\nu}^{\dagger}Y_{\nu})+(Y_e^{\dagger}Y_e)]
\right\}, 
\label{eq:RGEmNu} 
\end{align}
at the scale between 
$M_X$ and the typical mass scale of right-handed neutrinos $\overline{M}_R$
aside from the terms proportional to the unit matrix\cite{rge}.
In the region below $\overline{M}_{R}$, we use Eq.~\eqref{eq:RGEmNu}
but without the terms for $Y_{\nu}^{\dagger} Y_{\nu}$.    
The Yukawa matrix for charged leptons is given by 
$Y_{e} \simeq \text{diag}(0,0,y_{\tau})$ with 
$y_{\tau} = \sqrt{2} m_{\tau}/ (v \cos\beta)$.  
We concentrate on the case where
$Y_{\nu}^{\dagger}Y_{\nu}=\mathrm{diag}(y_1^2,y_2^2,y_3^2)$ for
simplicity.
We neglect the case in which the 2-3 element may affect on 
the physics as discussed in Ref.~\cite{ellis-hisano-raidal-shimizu}.
The solution of Eq.~\eqref{eq:RGEmNu} can be expressed 
as\cite{mst2,rge-sol}
\begin{align}
m_{\nu}(m_Z^{}) \simeq m_{\nu}(M_X)+Km_{\nu}(M_X)+m_{\nu}(M_X)K\;,
\label{solution-rge}
\end{align}
where $K = \text{diag}(\epsilon_{e}, 0, \epsilon_{\tau})$.
The $\epsilon_e$ and $\epsilon_{\tau}$ are given by\cite{st}
\begin{align}
\epsilon_e=&\frac{y_1^2-y_2^2}{16\pi^2}\ln \frac{M_X}{\overline{M}_{R}}\;,
\nonumber\\
\epsilon_{\tau}=&\frac{y_3^2-y_2^2}{16\pi^2}\ln \frac{M_X}{\overline{M}_R}
+\frac{y_{\tau}^2}{16\pi^2}\ln\frac{M_X}{m_Z^{}}\;.
\label{eps_y}
\end{align}

The experimental value $\theta_{\odot}$ can be reproduced from the bi-maximal
mixing solution at $M_{X}$ by the running effects.
The solar mixing angle $\theta_{\odot}$ is given in terms of
$\epsilon_{e,\tau}$, $\alpha_{0}$ and $m_{1}$ by
\begin{align}
\tan^2\theta_{\odot}=
\frac{1+2(\epsilon_{\tau}-2\epsilon_e)\cos^2(\alpha_0/2) (m_1^2/\Delta m_{\odot}^2)}
{1-2(\epsilon_{\tau}-2\epsilon_e)\cos^2(\alpha_0/2) (m_1^2/\Delta m_{\odot}^2)}\;,
\label{eq:tanSqthetaSol}
\end{align}
where $\Delta m_{\odot}^2$ is the experimental value 
for the mass-squared difference of solar neutrinos.
From Eq.~\eqref{eq:tanSqthetaSol}
the allowed region of $\alpha_{0}$ is obtained;
$\cos 2\theta_{\odot}\leq |\cos (\alpha_0/2)|$. 
In addition, the following condition among the Yukawa coupling constants is
found;
\begin{align}
2y_1^2>y_2^2+y_3^2+y_{\tau}^2.
\label{cond_yi2}
\end{align}
There are two possibilities under the condition Eq.~\eqref{cond_yi2} 
for the pattern of neutrino Yukawa couplings, i.e.,
the hierarchical case ($y_{3}^{2} < y_{2}^{2}  < y_{1}^{2}$) 
and the quasi-degenerate case ($y_{3}^{2} \simeq y_{2}^{2} < y_{1}^{2} $). 
As the order of $m_{Di}^{}$ is defined as 
$m_{D1}^{} \leq m_{D2}^{} \leq m_{D3}^{}$, 
$y_{i}$ $(i=1,2,3)$ are assigned as
\begin{align}
y_1= \frac{\sqrt{2} m_{D3}^{}}{v\sin\beta}, \quad 
y_2= \frac{\sqrt{2} m_{D2}^{}}{v\sin\beta}, \quad 
y_3= \frac{\sqrt{2} m_{D1}^{}}{v\sin\beta}\;,
\end{align}
which correspond that $V_L$ in Eq.~(\ref{vr-dd-vl}) is given by $P_{ex}$ in 
Eq.~\eqref{pex}.
\section{Derivations for $V_R$ and $M_i$}
\numberwithin{equation}{section}

We rotate $\widetilde{M}_R^{-1}$ 
in Eq.~(\ref{mrinv}) 
by 
$P_{ex}O_B$, and find
\begin{align}
(P_{ex}O_B)^T \widetilde{M}_R^{-1} P_{ex}O_B\simeq
\frac{m}{4m_{D1}^2}
\begin{pmatrix}
\left[(1+r)^2+(1-r)^2e^{i\alpha_0}\right]&
2(1-r^2)\cos\frac{\alpha_0}{2} e^{i\frac{\alpha_0}{2}}&0\\
2(1-r^2)\cos\frac{\alpha_0}{2} e^{i\frac{\alpha_0}{2}}&
\left[(1-r)^2+(1+r)^2e^{i\alpha_0}\right]&0\\
0&0&4 e^{i\beta_0}
\end{pmatrix}.
\end{align}
This matrix can be diagonalized by using the unitary matrix
\begin{align}
V_x=\begin{pmatrix}
\frac{1}{\sqrt{2}}             &\frac{1}{\sqrt{2}}e^{i\zeta}&0\\
-\frac{1}{\sqrt{2}}e^{-i\zeta} &\frac{1}{\sqrt{2}}          &0\\
0                              &0                           &1
\end{pmatrix}\;,
\end{align}
where $\zeta$ is defined in Eq.~\eqref{eq:defzeta}.
Consequently, we obtain 
\begin{align}
 V_R=P_{ex}O_BV_xP_{ex}^T
\mathrm{diag}(e^{-i\frac{\beta_0}{2}},e^{-i\frac{\alpha_0/2+\zeta}{2}},
e^{-i\frac{\alpha_0/2-\zeta}{2}}).
\end{align}
The explicit form is shown in Eq.~\eqref{eq_vr},
and those for $M_i$ are given in Eq.~(\ref{Mi-expression}).

\newpage



\end{document}